%% file: ParallelPerm.tex
\RequirePackage{rotating}
\documentclass[10pt]{iopart}

\usepackage{iopams,amsfonts,amssymb,amsthm} 

\pdfoutput=1
\usepackage{etex}
\usepackage{cite}
\usepackage{latexsym}
\usepackage{rawfonts}
\usepackage{graphicx}
\usepackage[usenames,dvipsnames,svgnames]{xcolor}
\usepackage{bbm}
\usepackage{latexsym}
\usepackage{multirow}
\usepackage{rotating}
\usepackage{lscape}
\usepackage{graphicx} 
\usepackage{subfigure}
\usepackage{xcolor}
\usepackage{fancybox}
\usepackage{ifmtarg}
\usepackage{fancyhdr}
\usepackage{wrapfig}

\input prepictex
\input pictex
\input postpictex

\usepackage{times}

\usepackage[font=footnotesize,labelfont=sf]{caption}

\include{macros}

\include{macros-pictex}

\begin{document}

\title[]{Parallel PERM}

\author{S Campbell$^1$ \& EJ Janse van Rensburg$^1$}

\address{\sf$^1$Department of Mathematics and Statistics, 
York University, Toronto, Ontario M3J~1P3, Canada\\}
\ead{rensburg@yorku.ca}
\vspace{10pt}
\begin{indented}
\item[]\today
\end{indented}

\begin{abstract}
We develop and implement a parallel flatPERM algorithm \cite{G97,PK04}
with mutually interacting parallel flatPERM sequences and use it to 
sample self-avoiding walks in 2 and 3 dimensions.  Our data show that 
the parallel implementation accelerates the convergence of the 
flatPERM algorithm.  Moreover, increasing the number of interacting flatPERM
sequences (rather than running longer simulations) improves the rate of
convergence.  This suggests that a more efficient implementation 
of flatPERM will be a massively parallel implementation, rather than long 
simulations of one, or a few parallel sequences.  We also use the algorithm 
to estimate the growth constant of the self-avoiding walk in two and in 
three dimensions using simulations over 12 parallel sequences.  Our 
best results are
\[
\mu_d = \cases{
2.6381585(1), & \hbox{if $d=2$}; \cr
4.684039(1), & \hbox{if $d=3$}.
} 
\]
\end{abstract}

%
\vspace{2pc}

\submitto{\JPA}

%
%

\section{Introduction}

The Rosenbluth algorithm \cite{HM54,RR55} samples self-avoiding walks  
by recursively appending steps at the end of the walk.  Since the sampling is not
uniform, the algorithm continuously updates a weight function which is used 
for determining averages of observables with respect to the uniform distribution 
over walks of length $n$ from the origin.  

More precisely, let $\mathS$ be the state space of self-avoiding walks from 
the origin $\vec{0}$ in the $d$-dimensional hypercubic lattice $\IntZ^d$, 
and denote the walk composed of the single vertex $\vec{0}$ and of 
length $0$ by $\emptyset$.  Suppose a walk $\omega_n$ of length $n$ 
steps have been grown recursively, then append the next step as follows:  
Let $a_+(\omega_n)$ be the number of possible (steps) edges incident with
the end-point of the $\omega_n$ which may be appended to $\omega_n$ 
to get a walk $\omega_{n+1}$ of length $n+1$.  Choose one of these 
edges uniformly, and append it to $\omega_n$ to obtain $\omega_{n+1}$.  
Recursive implementation of this generates a sequence (or chain) of walks 
$\LA \emptyset, \omega_1, \omega_2 , \ldots,\omega_n, \ldots \RA$, which we
shall also call a \textit{chain} (realised by the algorithm), and where $\omega_n$
is a \textit{prefix} of $\omega_{n+1}$. 

Since $a_+ (\emptyset) = 2d$, the probability of adding one step to grow the
walk $\omega_1$ of length one from $\emptyset$ is 
$\Pr(\emptyset\to\omega_1) = \Sfrac{1}{2d}$.  More generally, the probability 
of obtaining a walk $\omega_{n+1}$ of length $n+1$ from a walk $\omega_{n}$ 
of length $n$ is $\Pr(\omega\to\omega_{n+1}) 
= \Sfrac{1}{a_+(\omega_n)} \, \Pr(\omega_n)$.  

The probability of a particular sequence of walks
$S_n = \LA \emptyset, \omega_1, \omega_2 , \ldots,\omega_n\RA$ 
being realised by the algorithm is
\begin{equation}
\Pr (S_n) = \prod_{j=1}^n \Pr(\omega_{j-1} \to\omega_j) = \prod_{k=0}^{n-1}
 \Sfrac{1}{a_+(\omega_k)}
\label{eqn1}   
\end{equation}
where $\omega_0 = \emptyset$.

The \textit{weight of the sequence $S_n$} is defined by
\begin{equation}
W(S_n) = \Sfrac{1}{\footnotesize{\Pr(S_n)}} = \prod_{k=0}^{n-1} a_+(\omega_k).
\label{eqn2}   
\end{equation}

The function $W(S_n)$ is an observable, and its exact value for walks of length 
$n$, computed over all sequences of walks $S$ of length $|S|=n$ is 
\begin{equation} 
W_n = \sum_{S:|S|=n} \Pr(S)\, W(S) =  \sum_{S:|S|=n} \LB \Pr (S) \times \Sfrac{1}{\footnotesize{\Pr(S)}} \RB
= \sum_{S:|S|=n} {\LARGE 1} = c_n,
\label{eqn3}  
\end{equation} 
where $c_n$ is the number of walks of length $n$ from the origin and 
since each sequence $S$ of length $n$ ends in exactly one unique walk.  
Equation \Ref{eqn3} is the \textit{Rosenbluth counting theorem}.   
Estimating $W_n$ using the algorithm gives approximations of $c_n$, 
so that the Rosenbluth algorithm is an approximate enumeration algorithm.  
Since the algorithm grows walks recursively, there is a non-zero probability 
that a growing walk $\omega_n$ can be trapped (this occurs when $a_+(\omega_n)=0$) in which case $\Pr(\omega_n\to\omega^\prime)=0$.  
Any realised sequence or a chain which grow to include the trapped state 
$\omega_n$ is terminated, and the (hypothetical) subsequent states 
following $\omega_n$ are assigned the default weight zero.  Since any 
self-avoiding walk of length $n$ can be grown by the algorithm, this
algorithm is irreducible.  

Implementation of the algorithm to grow walks of length $n$
gives a set of independently grown walks of length $n$ denoted by 
$\{ \sigma_1, \sigma_2, \ldots, \sigma_M\}$ with weights $W_n(\sigma_j)$.  Since there
are trapped states, for some $\sigma_k$ it may be the case that $W_n(\sigma_k) = 0$.
The \textit{sample average} of $W_n(\sigma)$ is
\begin{equation}
\LH W_n \RH_M^{sample} 
= \sfrac{1}{M} \sum_{i=1}^M W_n (\sigmas{i})
= \Sfrac{\sum_{i=1}^M W_n (\sigmas{i})}{\sum_{i=1}^M W_0 (\sigmas{i})},
\label{eqn41AMC}  
\end{equation}
since $W_0 (\sigmas{i})=1$ and 
where $M$ is the number of sequences started by the algorithm (or more
accurately, the number of times the sequence passes through the empty walk and
restarts the sampling of a new walk), and $W_n(\sigmas{i})$
is the weight of the state $\sigmas{i}$.  By the strong law of large numbers
one expects that $\LH W_n \RH_M^{sample}$ converges to $W_n = c_n$ 
as $M\to\infty$ (see equation \Ref{eqn3}).

The estimator $\LH\C{O}\RH_M^{est}$ over a set of $M$ walks realised by the 
algorithm for the (canonical) average of an observable $\C{O}(\omega)$ over 
the uniform distribution of self-avoiding walks of length $n$ can computed using 
a ratio estimator:
\begin{equation}
\LH\C{O}\RH_M^{est} 
= \frac{\sum_{i=1}^M W_i (\sigmas{i})\thin \C{O}(\sigmas{i})}{
      \sum_{i=1}^M W_i(\sigmas{i})}  
=\frac{\LH W \C{O} \RH_M^{sample} }{\LH W \RH_M^{sample} } .
\label{eqn42MC}  
\end{equation}
As $M\to\infty$, then $\LH\C{O}\RH_M^{est} \to \LA \C{O} \RA$.

In this paper the feasibility of a parallel implementation of algorithms
based on Rosenbluth sampling (namely the PERM and flatPERM algorithms)
is considered.  The increasing parallel architecture of modern computers 
suggests that future improvements in performance will be obtained by 
implementing parallel versions of these algorithms, and such implementations 
may also bring improvements in convergence in the same way that gains 
were made by the introduction of multiple chains in parallel in Metropolis 
Monte Carlo methods \cite{TJvROW96}.  In the next section we briefly 
review PERM and flatPERM, and in section 3 we explain a parallel 
implementation of these algorithms.  This implementation is simple, and 
proceeds by seeding multiple PERM sequences in parallel (one per 
thread or CPU) and then collecting and sharing data between all the 
sequences as they evolve in real time.  

We test the parallel implementation and its performance in a variety of 
ways in section 3, including estimating $c_{10,000}$ (the number of 
self-avoiding walks of length $10,000$), a total absolute error for
simulations of walks up to length $10,\!000$, and estimating the 
least squares error and growth constant $\mu_d$ and entropic exponent
$\gamma$ for self-avoiding walks.  In section 4 we conclude the paper with 
a few final observations.

\section{PERM and flatPERM}

The Rosenbluth algorithm samples walks of moderate lengths (say
up to length $100$) very efficiently, but the attrition of walks due to 
trapped conformations in low dimensions, and the increasing dispersion 
of weights over a wide range of orders of magnitude as walks grow in 
length, quickly degrade estimators as $n$ increases (see equation 
\Ref{eqn42MC}).  As a result, alterations to the algorithm to compensate 
for the dispersion of weights, and attrition of walks, have been 
introduced.  These are variance reduction methods and they have
greatly improved the performance of the Rosenbluth algorithm.

The first variance reduction method is due to Meirovitch \cite{M85}, and 
is called the \textit{scanning method}.  Its implementation is not difficult, and 
it greatly improves the efficiency of Rosenbluth sampling by both dealing 
with the dispersion of weights and with attrition of walks due to trapped
conformations.  The second variance reduction method is due to 
Grassberger \cite{G97} (PERM), and a variant of this due to Prellberg and
Krawczyk \cite{PK04} (flatPERM) samples asymptotically over flat 
histograms over state space (flatPERM is also an example of rare event
sampling).

The PERM and flatPERM implementation of the Rosenbluth algorithm are based on
ideas of pruning and enrichment of states with low and high weights 
respectively \cite{WE59,FS79}.  These implementations were also generalised in
the flatGARM algorithm which is a more general algorithm based on Rosenbluth style 
sampling \cite{RJvR08}.

Suppose that a walk of length $n$ was grown using the Rosenbluth algorithm
by appending steps starting at the empty walk along a sequence
$\LA \omegas0,\omegas1,\ldots,\omegas{n}\RA$ (where $\omegas0 =\emptyset$).
The weight of state $\omegas{k}$ is denoted by $W(\omegas{k})$ and is
given by equation \Ref{eqn2} where 
$S_k = \LA \omegas0,\omegas1,\ldots,\omegas{k}\RA$ so that 
$W(\omegas{k}) \equiv W(S_k)$.  

Introduce a cut-off $T_k$ on $W(\omegas{k})$ for walks of length $k$.  If 
$W(\omegas{k} )> T_k$, then \textit{enrich} $\omegas{k}$ in $S_k$ by adding $M$ 
copies of $\omegas{k}$ to $S_k$ and by reducing (dividing) $W(\omegas{k})$ by a 
factor of $M$.  The algorithm then continues to grow $M$ walks from 
$\omegas{k}$ independently with reduced weights, in each case continually 
enriching states if their weights similarly exceed the cut-off $T_k$. This 
enrichment and weight reduction of states with large weights have the 
effect of reducing the dispersion of weights systematically.  Enriching states 
also does not disturb the sample average of observables.

A state $\omegas{k}$ with a small weight can be pruned by removing it from
$S$ and assigning it zero weight.  This is implemented by introducing a lower 
cut-off $t_k$ at length $k$ on $W(\omegas{k})$.  If $W(\omegas{k}) < t_k$, 
then the walk is \textit{pruned} with probability $1\minus  \sfrac{1}{q}$ 
where $q$ is a parameter of the algorithm.  If the walk is 
not pruned (with probability $\sfrac{1}{q}$), then its weight 
is increased by a factor of $q$. Similarly to enrichment, pruning a state
with low weight does not disturb sample averages.

The dispersion of the weights $W(S_n)$ in PERM may be further reduced
by taking the cut-offs in its implementation to be equal ($t_k = T_k$)
and then to continually enrichment and prune states exceeding or falling
below the cut-off.  This is implemented as follows: Let $[W_k]_M^{sample}$ 
be the running average of the weights of walks of length $k$ after $M$ 
sequences were realised by the algorithm.  If the walk $\omegas{k}$ 
in the $M$-th sequence has PERM weight $W(\omegas{k})$, then compute 
the ratio 
\begin{equation}
r = \frac{W(\omegas{k})}{[W_k]_M^{sample}},
\label{eqnr}      
\end{equation}
and where the weight $W(\omegas{k})$ is also included in the calculation
of $[W_k]_M^{sample}$.  The value of $[W_k]_M^{sample}$ serves
as a cut-off.  If $r>1$ then the weight $W(\omegas{k})$ exceeds its
expected value, and the state may be enriched, and if $r<1$ then the
state has lower than expected weight, and may be pruned. 

If $r\geq 1$ then the state could be enriched.  Compute probability 
$p={\lcl r\rcl} \minus  r$ and put $c=\lfl r\rfl$ with probability $p$ and 
with default $c = \lcl r \rcl$.  Place $c$ copies of $\omegas{k}$ in the 
sequence, each with reduced weight $\sfrac{1}{c} W(\omegas{k})$.  
Continue to grow the sequence from each of these states independently, 
and at each iteration, determine $r$ as above.

If $r<1$ then $W(\omegas{k})$ is smaller than expected.  Prune it with 
probability $1-r$.  If it is not pruned, then increase $W(\omegas{k})$ by
multiplying it with $\sfrac{1}{r}$.  

In flatPERM simulations the running average $[W_k]_M^{sample}$ of
weights is initially poor but improves quickly, and the sampling stabilizes to
flat histogram sampling.   There are very low attrition of sequences,
and the variance reduction in flatPERM gives a quickly convergent algorithm 
sampling over weights in a narrow range.

\begin{figure}
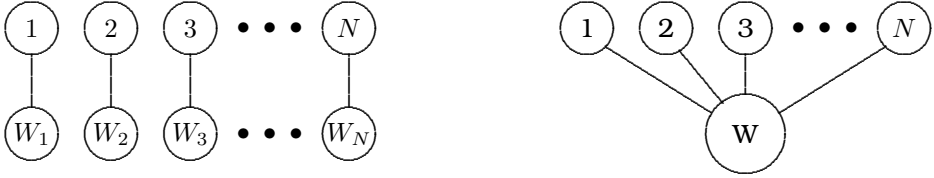

\beginpicture
\setplotarea x from -30 to 400, y from -30 to 60

\multiput
{\beginpicture \circulararc 360 degrees from 10 0 center at 0 0 \endpicture}
at 0 40 30 40 60 40 120 40  / 
\multiput
{\beginpicture \circulararc 360 degrees from 10 0 center at 0 0 \endpicture}
at 0 0 30 0 60 0 120 0  / 
\multiput {$\bullet$} at 80 40 90 40 100 40 /
\multiput {$\bullet$} at 80 0 90 0 100 0 /
\multiput {\beginpicture \plot 0 10 0 30 / \endpicture} at 0 0 30 0 60 0 120 0 /

\put {$1$} at 0 40 
\put {$2$} at 30 40 
\put {$3$} at 60 40 
\put {$N$} at 120 40

\put {$W_1$} at 0 0 
\put {$W_2$} at 30 0 
\put {$W_3$} at 60 0 
\put {$W_N$} at 120 0

\setcoordinatesystem units <1pt,1pt> point at 20 0
\multiput
{\beginpicture \circulararc 360 degrees from 10 0 center at 0 0 \endpicture}
at 250 40 280 40 310 40 370 40  / 
\multiput
{\beginpicture \circulararc 360 degrees from 15 0 center at 0 0 \endpicture}
at 310 0   / 

\setcoordinatesystem units <1pt,1pt> point at 40 0
\multiput {$\bullet$} at 330 40 340 40 350 40 /
\plot 297 8 257 33 /
\plot 301.5  11.5  285 31.5 /
\plot 310 15  310 30 /
\plot 323 8  363 33 /

\put {1} at 250 40 
\put {2} at 280 40 
\put {3} at 310 40 
\put {$N$} at 370 40 

\put {\scalebox{1.0}{W}} at 310 0

\endpicture
\caption{Parallel implementation of PERM.  
Algorithm 1:  On the left $N$ independent PERM sequences sample independent 
walks with sample averages of weights $[W_k]^{sample}_M$ independently
computed for each sequence.  Enrichment and pruning of states are 
independently done in each sequence.  
Algorithm 2:  On the right the $N$ sequences are interacting with each other 
by sharing collected weights (running averages of weights) in a single data 
location $W$.  Enrichment and pruning of states in each sequence are done by 
comparing weights with the sample average of the weights computed over 
data collected over the $N$ sequences, including the data from partially 
completed sequences.  In both algorithms each sequence is realised in its 
own CPU (or thread) in the computer and runs simultaneously with other 
sequences.  While running averages of weights are quarantined in each 
realised sequence in Algorithm 1, in Algorithm 2 the weights are pooled into 
a shared data structure $W$ while each sequence is sampling along its own 
thread while accessing the pooled data to calculate enrichment and pruning 
parameters $p$ and $r$.  The CPU times of Algorithms 1 and 2 are virtually 
the same for the same number of parallel sequences and iterations (or started
walks).}
\label{figureA}
\end{figure}

\section{Parallel PERM}

Two parallel implementations of the PERM algorithm are shown in figure \ref{figureA}.  
We consider them in turn

\vspace{5mm}
\noindent{\underbar{Algorithm 1:}}
On the left $N$ independent realisations of the algorithm (1 per CPU or per thread)
are initiated.  The $\ell$-th realisation calculates a sample average of weights
$[W_k]^{sample}_{M,\ell}$ for $\ell=1,2,\ldots,N$ given by equation \Ref{eqn41AMC}.
The average over the $N$ parallel sequences is
\begin{equation}
[W_k]^{(\hbox{\fns alg 1})}_{M,N} = \Sfrac{1}{N} \sum_{\ell=1}^N [W_k]^{sample}_{M,\ell}
\end{equation}
where $M$ is the length of each sequence, and $k$ is the length of the walk.  By
the strong law of large numbers, $[W_k]^{parallel}_{M,N}$ converges to $c_k$
if $M\to\infty$.  The convergence is accelerated if $N$ is increased (that is, when more
sequences are initiated in parallel).

\vspace{5mm}
\noindent{\underbar{Algorithm 2:}}
An integrated parallel implementation of $N$ realised PERM sequences sharing
data is shown on the right in figure \ref{figureA}.  As opposed to the implementation
on the left, this is a true parallel implementation in that the $N$ parallel sequences
are not independent of each other, but communicate continuously by accessing 
data generated by all other sequences.  These shared data are used to determine
the enrichment and pruning in each of the parallel sequences, and each sequence
is continuously updating the shared data as it progresses.  The average weights
in this case are denoted by $[W_k]^{(\hbox{\fns alg 2})}_{M,N}$ and it is 
computed by using equation \Ref{eqn41AMC} (where $M$ is now the total
number of passes of all sequence through the trivial walk of length zero).

\vspace{5mm}
The flatPERM algorithm was implemented using both Algorithm 1 and Algorithm 2
to sample along multiple sequences.   In both implementations the algorithms were 
coded in $C$ with open-mp protocols \cite{Open-MP} to access CPUs and to place 
one PERM sequence per thread.  These algorithms were run on a desktop workstation
and tested for convergence in various ways.  Our results are shown below.

\subsection{Estimating $c_{10,000}$}
In table \ref{tableA} the results for simulations with two sequences are shown.  
The number of started walks (iterations) is given in the top row (this is the 
total number of walks generated -- since there are two sequences the number 
of walks per CPU is given by the powers of $10$ in each case).   Estimates of 
$\log c_{10,000}$ were made by calculating the average weights of the 
realised sequences.  In Algorithm 1 an estimate was obtained for each 
independent sequence, and the best estimate was calculated by taking the 
geometric average of the estimates from each sequence.  In the case of 
Algorithm 2 there is only one set of data collected over all sequences, and
the estimate of $\log c_{10,000}$ was obtained in this case by estimating the
weight over all the pooled data.  These estimates are listed in the third row of
table \ref{tableA}.

\begin{table}[h!]
\caption{$\log c_{10,000}$ estimated using $2$ sequences}
\centering
\begin{tabular}{|l||r|r|r|r|r|r|}
\hline
\#Walks & $2\times1$ & $2\times10$ & $2\times10^2$ & $2\times10^3$ 
             & $2\times10^4$ & $2\times10^5$ \\
\hline
Alg. 1 & $9226.07$ & $9601.69$ & $9658.76$ & $9701.30$ & $9704.15$ & $9704.11$ \\
Alg. 2 & $9491.40$ & $9644.84$ & $9702.71$ & $9705.09$ & $9703.88$ & $9704.13$ \\
\hline
\end{tabular}
\label{tableA}
\end{table}

The estimate by Algorithm 1 settles down by $2\times 10^4$ iterations 
(started walks) at a value close to $9704$.  By increasing the number of iterations
by a factor of $10$ each along the columns of the first row, the estimate in the 
second row is seen to increase as the simulation proceeds before it levels off.  A similar 
pattern is seen for Algorithm 2 -- however, it levels off close to $9704$ already by
$10^2$ walks, and more definitely by $10^3$.  This is a factor of
$10$ faster than the convergence seen in Algorithm 1.  These data and results seem
to imply that Algorithm 2 gives a gain of a factor of about $10$ in convergence of the
approximate estimates of $c_{10,000}$ in simulations which sample walks
up to length $10^4$.  For comparison, a very long simulation using flatPERM
($1.6\times 10^9$ iterations) gives
the estimate
\begin{equation}
\log c_{10,000} = 9704.14\ldots
\label{eqn8} 
\end{equation}
in the square lattice.

Similar results are seen when more sequences are used in the simulations.  In table
\ref{tableB} results similar to those in table \ref{tableA} are shown,  but now for walks sampled
using $12$ sequences.  These results again level off with increasing number of walks, and 
again a large gain is seen for Algorithm 2.  By $12\times 10$ walks the estimate is within
$7$ of the value in equation \Ref{eqn8}, while for Algorithm 1 it is still about
$65$ below.

\begin{table}[h!]
\caption{$\log c_{10,000}$ estimated using $12$ sequences}
\centering
\begin{tabular}{|l||r|r|r|r|r|r|}
\hline
\#Walks & $12\times1$ & $12\times10$ & $12\times10^2$ & $12\times10^3$ 
             & $12\times10^4$ & $12\times10^5$ \\
\hline
Alg. 1 & $9273.79$ & $9658.79$ & $9692.63$ & $9703.54$ & $9704.38$ & $9704.11$ \\
Alg. 2 & $9652.09$ & $9697.79$ & $9703.23$ & $9704.04$ & $9704.16$ & $9704.18$ \\
\hline
\end{tabular}
\label{tableB}
\end{table}

The results in tables \ref{tableA} and \ref{tableB}  show that Algorithm 2
outperforms Algorithm 1 substantially in particular at the initial stage of the algorithm
(after a few walks have been sampled).  Convergence of Algorithm 1 appears 
to occur when the number of iterations (started walks) approach about 
$12\times 10^4$ while Algorithm 2 is already close to its target 
after $12\times 10^2$ iterations.  This shows a substantial 
increase in the rate of convergence of flatPERM with the introduction of coupling 
between sequences as proposed in Algorithm 2.  In addition, the results for Algorithm 2 in 
tables \ref{tableA} and \ref{tableB} show that increasing the number of sequences 
from two to twelve improves the results for lower number of walks, as expected.  
A similar gain is seen for Algorithm 1, but not to the same degree.

\subsection{Total absolute error}

We define the total absolute error per unit length of $\log c_n$ by
\begin{equation}
T_M = \sfrac{1}{M} \sum_{n=1}^{M} \LV  \log c_n^{best} - \log c_n^{est} \RV
\label{eqn9}
\end{equation}
where $c_n^{best}$ is the best estimates of $c_n$ and $c_n^{est}$ is the
estimate for $c_n$ obtained by either algorithm 1 or algorithm 2.  $M$ is
the maximum length of walks sampled by the algorithms.

Good estimates for $c_n^{best}$ are obtained from a very long (ordinary) 
flatPERM simulation of $1.6\times 10^9$ iterations.  For $c_n^{est}$ the 
average weights $[W_k]^{(\hbox{\fns alg 1})}_{M,N}$ and
$[W_k]^{(\hbox{\fns alg 2})}_{M,N}$ are used respectively, and in each case 
$T_M$ will be an estimate of the total deviation per unit length of the estimates 
from the best values $\log c_n^{best}$.  For example, determining $T_M$ from one
sequence growing a single walk with the flatPERM algorithm gives
a total absolute error per unit length of $239.33$ for walks of length up to
$M=10,000$.   

The results are shown in table \ref{tableC} for both
algorithms and for lengths of walks up to $10^4$.  The first column gives
the number of started walks \textit{per sequence} for each algorithm ($S$).
The columns under Algorithm 1 shows $T_{10,000}$ as measured using
equation \Ref{eqn9}.  For example, a simulation of Algorithm 1 using 2 sequences
for $1$ started walk each gives the total absolute error $231.79$, as seen in
the column $N=2$ under algorithm 1, while using two sequences
in parallel in Algorithm 2 gives $113.17$, a significant reduction as seen
in the column $N=2$ under Algorithm 2, especially at lower numbers of
started walks.

\begin{table}[t!]
\caption{$T_{10,000}$}
\centering
\begin{tabular}{|l||r|r|r|r|r||r|r|r|r|}
\hline
 & \multicolumn{5}{c||}{Algorithm 1} &  \multicolumn{4}{c|}{Algorithm 2} \\
\hline
S & $N=1$ & $N=2$ & $N=3$ & $N=6$ & $N=12$ & 
$N=2$ & $N=3$ & $N=6$ & $N=12$  \\
\hline
$1$ &  $239.33$ & $231.79$ & $224.73$ & $213.67$ & $200.88$ & $113.17$ & $108.71$ & $44.81$ & $20.38$ \\
$10$ & $69.34$ & $47.22$ & $47.53$ & $48.19$ & $22.26$ & $32.49$ & $13.58$ & $8.01$ & $2.89$  \\
$10^2$ & $30.72$ & $23.61$ & $14.39$ & $6.65$ & $5.59$ & $1.120$ & $0.716$ & $0.500$ & $0.339$  \\
$10^3$ & $3.217$ & $1.284$ & $1.256$ & $0.696$ & $0.220$ & $0.620$ & $0.812$ & $0.393$ & $0.280$ \\
$10^4$ & $0.383$ & $0.197$ & $0.0920$ & $0.271$ & $0.145$ & $0.190$ & $0.0767$ & $0.0657$ & $0.0459$ \\ 
$10^5$ & $0.0446$ & $0.0387$ & $0.0466$ & $0.0481$ & $0.0310$ & $0.0519$ & $0.0928$ & $0.0188$ & $0.0144$ \\
$10^6$ & $0.0271$ & $0.0106$ & $0.0063$ & $0.0047$ & $0.0036$ & $0.0098$ & $0.0094$ & $0.0057$ & $0.0111$ \\
\hline
\end{tabular}
\label{tableC}
\end{table}

The results in table \ref{tableC} show that for each algorithm there is improved 
performance down each column (that is, increasing the number of started walks 
per sequence), and along each row (increasing the number of sequences and thus 
the total number of started walks).  Since $T_M$ is the average of 
$| \log (c_n^{est}/c_n^{best}) |$ over all values of $n \leq M$, its best value is zero, 
and large values are indicative of poor convergence of the algorithm.  The data 
suggest that convergence is good when there are $10^6$ started walks in each 
squence, regardless of the number of independent or parallel sequences.   The 
data also shows far superior performance for Algorithm 2, even at modest 
values of the number of started walks per sequence.  For example, for two sequences 
at just 100 walks per sequence, $T_{10,000}$ is reduced from $23.61$ to $1.120$ if 
the sequences are coupled as in Algorithm 2. Similar results are seen as the number of sequences are increased in Algorithm 2.

\subsection{Estimating $\mu$ and $\gamma$}

The growth constant $\mu_d$ of self-avoiding walks in the $d$-dimensional hypercubic 
lattice is defined by the limit \cite{HM54}
\begin{equation}
\lim_{n\to\infty} \Sfrac{1}{n} \log c_n = \mu_d .
\end{equation}
It is also known that \cite{K63}
\begin{equation}
\lim_{n\to\infty} \Sfrac{c_{n+2}}{c_n} = \mu_d^2 .
\label{eqn11}   
\end{equation}
It is not known that the limit $\lim_{n\to\infty} (c_{n+1}/c_n)$ exists, but the above 
shows that $c_n = \mu_d^{n+o(n)}$.   The result in equation \Ref{eqn8} shows
that $\log \mu_d \approx 0.970$ in the square lattice.   The best numerical estimates of
$\mu_d$ in the square and cubic lattices are
\begin{equation}
\mu_d = \cases{
2.63815853035(2), & if $d=2$ \cite{CJ12}; \cr
4.684039931(27), & if $d=3$ \cite{C13}.
}
\label{12}
\end{equation}
Taking logarithms gives the best estimates
\begin{equation}
\log \mu_d = \cases{
0.970081147258(8), & if $d=2$ \cite{CJ12}; \cr
1.5441609707(58), & if $d=3$ \cite{C13}.
}
\label{13}
\end{equation}

There is numerical evidence that
\begin{equation}
c_n = C\,n^{\gamma-1} \, \mu_d^n\, (1+o(1))
\label{eqn13}   
\end{equation}
where $\gamma$ is the \textit{entropic exponent}.  In two dimensions the exact value
of $\gamma = \sfrac{43}{32}$ \cite{D86} while in three dimensions 
$\gamma=1.15698(34)$ \cite{S11}.

The efficiency of Algorithms 1 and 2 will be examined by calculating estimates of $\mu_d$
and $\gamma$ from our data, controlling for the number of sequences and increasing
the number of walks per sequence.  In order to estimate $\mu_d$, consider the ratio
\begin{equation}
\Sfrac{c_{n+1}}{c_n} = \mu_d\, \LB 1 + \Sfrac{1}{n} \RB^{\gamma-1} \, (1+o(1)) 
\end{equation}
inspired by equations \Ref{eqn11} and \Ref{eqn13}. Taking logarithms gives the model
\begin{equation}
\log \LB \Sfrac{c_{n+1}}{c_n} \RB \approx 
\log \mu_d + (\gamma-1)\log \LB 1 + \Sfrac{1}{n} \RB + \Sfrac{c}{n^2}
\label{eqn15}  
\end{equation}
where the last term is inserted as the first analytic correction.  A three parameter
linear least squares regression will give estimated values for $\mu_d$ and $\gamma$.
Improved estimates of $\gamma$ are obtained by fixing $\mu_d$ at its
best value in equation \Ref{13} and then using a two-parameter fit to
estimate $\gamma$.

The performance of the algorithms can also be examined by looking at the
level of noise in the estimates of $\log ( \Sfrac{c_{n+1}}{c_n} )$ as a function
of $n$. Since the correction terms in equation \Ref{eqn15} approach zero fast, 
these estimates should scatter in a band around the right hand side of 
equation \Ref{eqn15} and the width of the band will be a measure of how 
well converged the data are.

\begin{figure}[t!]
  \includegraphics[width=\linewidth]{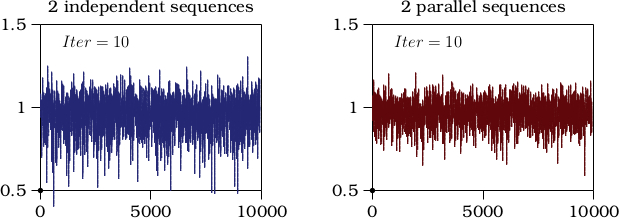}
  \caption{The variability of $\log (c_{n+1}/c_n)$ plotted against $n$ for
a simulation of  $2$ chains and $10$ iterations per chain.}
  \label{Figure2}
\end{figure}

In figure \ref{Figure2} these data are shown for 2 sequences with $10$ walks
generated by the Algorithms.  In the panel on the left the data are shown for
Algorithm 1, and on the right, for Algorithm 2.  The width of the band can be
estimated by computing the root of the least square error $E$ of a regression fitting
$\log ( \Sfrac{c_{n+1}}{c_n} )$ to the right hand side of equation \Ref{eqn15}.
In this case the results are $E=0.1129$ on the left, and $E=0.08186$ on the right,
confirming the perception that the band in the left panel is wider than the band
in the right panel.  In other words, the data obtained by Algorithm 2 
are more clustered to the regression line, than the data obtained by Algorithm 1.

\begin{figure}[t!]
  \includegraphics[width=\linewidth]{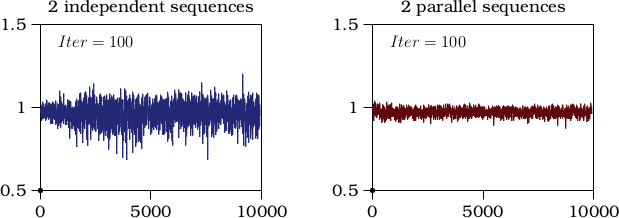}
  \caption{The variability of $\log (c_{n+1}/c_n)$ plotted against $n$ for
a simulation of  $2$ chains and $100$ iterations per chain.}
  \label{Figure3}
\end{figure}

\begin{figure}[t!]
  \includegraphics[width=\linewidth]{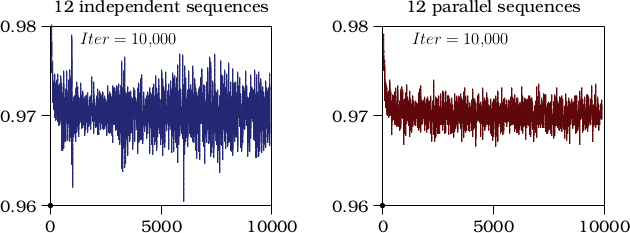}
  \caption{The variability of $\log (c_{n+1}/c_n)$ plotted against $n$ for
a simulation of  $12$ chains and $10,\!000$ iterations per chain.}
  \label{Figure4}
\end{figure}

Increasing the number of walks per sequence to $100$ gives the results in
figure \ref{Figure3}.  Both the bands are markedly narrower than in
figure \ref{Figure2}, and the values of $E$ confirm this, namely
$0.06548$ for the left panel, and $0.0206$ for the panel on the right.
This also supports a conclusion that the rate of convergence for Algorithm 2
is faster than that of Algoritm 1.  Another example, in this case for
$12$ sequences and $10^4$ started walks, are shown in figure
\ref{Figure4},  here the $E$ are $0.001951$ and $0.001091$, respectively, 
for the left and right panels.

We have calculated $E$ for all our data and the results
are shown in table \ref{table4}.    The notation is compacted so that
$\o{2}.1122 \equiv 0.001122$ (that is, the barred digit is the number
of zeros following the decimal point).  Notice that Algorithm 2 consistently has
smaller values for shorter runs, but that this advantage shrinks are longer
simulations are done. By $10^6$ started walks, the widths are largerly
the same.  This suggest that the acceleration of convergence due to the
parallel implementation in Algorithm 2 is best exploited by performing
shorter simulations of the parallel implementation, and then to combine 
the results of several independent simulations for final results.  In other
words, more parallel sequences, rather than longer simulations, is the key to 
quick convergence and good results, and massively parallel implementations
of Algorithm 2 may be the best approach.

\begin{table}[t!]
\caption{$E$ for Algorithms 1 and 2}
\centering
\begin{tabular}{|r||r|r|r|r|r||r|r|r|r|}
\hline
 & \multicolumn{5}{c||}{Algorithm 1} &  \multicolumn{4}{c|}{Algorithm 2} \\
\hline
N & $N=1$ & $N=2$ & $N=3$ & $N=6$ & $N=12$ & 
$N=2$ & $N=3$ & $N=6$ & $N=12$  \\
\hline
$1\;$ &  $0.2782$ & $0.2578$ & $0.2622$ & $0.2708$ & $0.2639$ & $0.1886$ & $0.1580$ & $0.1113$ & $\o{1}.8023$ \\
$10\;$ & $0.1363$ & $0.1129$ & $0.1129$ & $0.1129$ & $\o{1}.8501$ & $\o{1}.8186$ & $\o{1}.5643$ & $\o{1}.3870$ & $\o{1}.2556$  \\
$10^2$ & $\o{1}.8394$ & $\o{1}.6548$ & $\o{1}.4348$ & $\o{1}.3292$ & $\o{1}.2388$ & $\o{1}.2060$ & $\o{1}.1754$ & $\o{1}.1284$ & $\o{2}.8540$  \\
$10^3$ & $\o{1}.2783$ & $\o{1}.1103$ & $\o{1}.1077$ & $\o{1}.1112$ & $\o{2}.5158$ & $\o{2}.8695$ & $\o{2}.7670$ & $\o{2}.4156$ & $\o{2}.3038$  \\
$10^4$ & $\o{2}.3329$ & $\o{2}.2493$ & $\o{2}.2130$ & $\o{2}.2903$ & $\o{2}.1951$ & $\o{2}.2811$ & $\o{2}.1979$ & $\o{2}.1551$ & $\o{2}.1091$  \\
$10^5$ & $\o{2}.1244$ & $\o{3}.8238$ & $\o{3}.7084$ & $\o{3}.4961$ & $\o{3}.3755$ & $\o{3}.8673$ & $\o{3}.7317$ & $\o{3}.4936$ & $\o{3}.3511$  \\
$10^6$ & $\o{3}.3980$ & $\o{3}.2771$ & $\o{3}.2294$ & $\o{3}.1590$ & $\o{3}.1125$ & $\o{3}.2742$ & $\o{3}.2262$ & $\o{3}.1575$ & $\o{3}.1120$  \\
\hline
\multicolumn{10}{|l|}{Notation:  $\o{a}.b_1b_2b_3b_4 \equiv 0.b_1b_2b_3b_4\times 10^{-a}$} \\
\hline
\end{tabular}
\label{table4}
\end{table}

As a final test of our implementation we estimated the growth constants 
$\mu_2$ and $\mu_3$.  In the square lattice we performed two simulation 
of walks of lengths up to $50,\!000$.  The first simulation was stopped after 
a total of $105,\!685,\!556$ iterations (started walks) over $12$ parallel
sequences (or about $8,\!807,\!130$ per parallel sequence), 
and the second was run to completion at $120,\!000,\!000$ iterations (started 
walks) over $12$ parallel sequence (or $10,\!000,\!000$ per parallel
sequence).   A three parameter fit of equation 
\Ref{eqn15} to the weighted geometric average of the data for lengths 
$1\leq n \leq 50,\!000$ was used to determine an estimate of $\mu_2$.  This 
shows that
\begin{equation}
\log \mu_2 = 0.970081152
\label{eqn16}
\end{equation}  
compared to the estimate of $\log \mu_2 = 0.970081147258(8)$ by Clisby 
and Jensen \cite{CJ12},  showing that our result is within $5\times 10^{-9}$ 
from their more accurate estimate.  We do confirm the first 6 digits in the 
decimal expansion.

Next, we consider estimates of $\mu_3$ in the cubic lattice using more
extensive simulations in order to both determine the efficiency of the algorithm,
and to find good estimates of the growth constant.

In the cubic lattice we performed seven simulations of walks of lengths up 
to $50,\!000$ using Algorithm 2 with 12 parallel sequences, and discarded 
data for lengths over $49,\!900$ from our data due to boundary effects.  
The first simulation was of length $4,\!250,\!000$ iterations
per parallel sequence (for a total of $51,\!000,\!000$ iterations), and the 
remaining six simulations were each of length $2,\!125,\!000$ iterations 
per parallel sequence (for a total of $25,\!500,\!000$ iterations for each 
simulation).  These simulation give the $7$ estimates
\begin{equation}
\log \mu_3  = 
\left\{
\begin{array}{lll}
 1.5441604989 & 1.5441608584  & 1.5441603031  \\
 1.5441607748 & 1.5441611068  & 1.5441612779  \\
 1.5441608359 \\  
\end{array} \right.
\label{eqn17A}
\end{equation}
each stated to 10 decimal places.  A weighted average of these results give
$\log \mu_3=1.544160769$.  Rounding our result and comparing it to
the best estimate by Clisby \cite{C13}, namely 
$\log \mu_3 = 1.5441609707(58)$, show that we have verified six 
decimal places, namely
\begin{equation}
\log \mu_3 = 1.\underbar{544160}77 .
\label{eqn18}
\end{equation}
If, instead, the geometric averages over all the data in the seven simulations
are taken, and then analysed, we obtain the estimate
\begin{equation}
\log \mu_3 = 1.54416076 .
\label{eqn19}
\end{equation}
The total number of iterations, over all the simulations and sequences, 
is $204,\!000,\!000$.

The efficiency of algorithm 2 is best illustrated by performing shorter simulations,
and comparing the results to the above.  Simulations of walks to length $n=50,\!000$
were again performed, but now doing $200,\!000$ iterations per thread
along $12$ sequences (for a total of $2,\!400,\!000$ per simulation - each of
these simulation was just about $9.4$\% of the length of those leading to the
results in equation \Ref{eqn17A}).  Over $31$ simulations we did a total of 
$74,\!400,\!000$ iterations.  These simulations give the following results
which are comparable in accuracy to the results in equation \Ref{eqn17A} and
are
\begin{equation}
\hspace{-2cm}
\log \mu_3  = 
\left\{
\begin{array}{llll}
1.5441593027 & 1.5441599034 & 1.5441605413 & 1.5441604908 \\
1.5441611903  & 1.5441606073 & 1.5441603137 & 1.5441612410 \\
1.5441614305 & 1.5441604872 & 1.5441594056 & 1.5441608619 \\
1.5441599673 & 1.5441620647 & 1.5441622034 & 1.5441603728 \\
1.5441611795 & 1.5441607140 & 1.5441629101 & 1.5441609189 \\
1.5441591882 & 1.5441618451 & 1.5441618095 & 1.5441608069 \\
1.5441596354 & 1.5441606142 & 1.5441607758 & 1.5441595308 \\
1.5441597235 & 1.5441612369 & 1.5441619484 
\end{array} \right.
\label{21}
\end{equation}
with results stated to 10 decimal places.  Taking a simple average over these results 
gives $\log \mu_3=1.544160749$, showing that these data are converged.  This
result rounds to $1.54416075$ and is comparable to the result in equation 
\Ref{eqn18}.  If the geometric average over all the data in the $31$ simulations 
are taken and then $\log \mu_3$ is computed, then the result is again
\begin{equation}
\log \mu_3 = 1.54416075 ,
\label{22}
\end{equation}
as compared to equation \Ref{eqn19}.

The above results strongly suggest that shorter simulations using more
sequences in parallel for longer walks give superior performance, at least
when the aim is to estimate $\log \mu_d$ (our data also show that longer
simulations are needed to get good results for the entropic exponent
$\gamma$).

As a final test we sampled walks of length $n=200,\!000$.  Nine simulations
with $100,\!000$ started sequences per thread (for a total of $12 \times 
100,\!000 = 1,\!200,\!000$ iterations per simulation, or just half the number
of started sequences per thread leading to the results in equation \Ref{21}) 
give the following estimates 
\begin{equation}
\hspace{-1cm}
\log \mu_3 = 
\left\{
\begin{array}{llll}
1.5441618451 & 1.5441630316 & 1.5441630117 \cr
1.5441625924 & 1.5441626936 & 1.5441623028 \cr
1.5441628816 & 1.5441620305 & 1.5441633443
\end{array} \right.
\label{23}
\end{equation}
A simple average over the nine estimates gives $\log \mu_3 = 1.5441626371$.
This slightly exceeds the better estimate in equation \Ref{22}, but given
that these simulations were very short (each taking an average of just
16.5 hours CPU time).  The total number of iterations is $10,\!800,\!000$.  Since these
walks were also longer than those in equation \Ref{21}, a better estimate of
the exponent $\gamma$ is obtained, namely
\begin{equation}
\hspace{-1cm}
\gamma = 
\left\{
\begin{array}{llll}
1.1608790435 & 1.1572583036 & 1.1613650153 \cr
1.1543650430 & 1.1581152153 & 1.1578135872 \cr
1.1576672680 & 1.1513397803 & 1.1561657305
\end{array} \right.
\label{24}
\end{equation}
Taking a simple average gives $\gamma = 1.15722$, which compares well
with the result in reference \cite{S11}.


\section{Conclusions}

Our data clearly show that the parallel implementation of flatPERM using 
algorithm 2 (see figure \ref{figureA}) outperforms flatPERM as implemented
using algorithm 1.  Since algorithms 1 and 2 use the same computational
resources (for example, CPU time and number of threads) our approach
to analyse output in order to compare performance is a fair comparison
to determine the relative improvement seen in algorithm 2 over algorithm 1.

The improvement seen in algorithm 2 is in particular evident by the reduction
in the time it takes to see convergence after it is initialised. In addition,  there 
is also a noticeable improvement with increasing the number of parallel 
sequences in algorithm 2.  This is seen, for example, in table \ref{tableC} 
where there is an improvement with increasing number of sequences for 
low numbers of iterations.  Similar improvements are seen in tables 
\ref{tableA} and \ref{tableB}.  However, the reduction in noise with
the increasing number in sequences, in particular for algorithm 2, as 
shown in figures \ref{Figure2}, \ref{Figure3} and figure \ref{Figure4}, 
is more dramatic, and this is confirmed by the data in table \ref{table4} 
showing that algorithm 2 outperforms algorithm 1 in particular when each 
parallel sequence is shorter than about $10^4$ iterations (started walks). 

We also estimated the growth constant $\mu_d$ for walks using 
algorithm 2.  Our best results are
\begin{equation}
 \mu_d = \cases{
2.6381585(1), & \hbox{if $d=2$}; \cr
4.684039(1), & \hbox{if $d=3$},
} 
\label{bestmu}
\end{equation}
obtained by exponentiating the results in equations \Ref{eqn16} and 
\Ref{eqn19} and rounding it in $d=2$ to seven decimal places, and in 
$d=3$ to six decimal places. The result in $d=2$ is different by about 
$1\times 10^{-7}$ from the result in reference \cite{CJ12}, and that in 
$d=3$ by less than $1\times 10^{-6}$ from the result in reference \cite{C13}; 
see equation \Ref{12}.  The simulations leading to the results in 
equation \Ref{eqn17A} were all done on a single DELL Optiplex Desktop 
workstation, and the results in equations \Ref{21} and \Ref{23} were obtained by
submitting our programs to a Dell R340 node with 12 threads
(and 6 cores).

Using our data leading to the estimates in equation \Ref{eqn17A}
we also estimated the entropic exponent $\gamma$ (see equation 
\Ref{eqn15}).  This was done by using the three parameter model
in equation \Ref{eqn15} with a minimum cut-off $n_{min}$ for $n$ 
(that is, for $n\geq n_{min}$).  By extrapolating the results against $n_{min}$ 
our best estimates are
\begin{equation}
\gamma = \cases{
1.34416(42) & \hbox{if $d=2$}; \cr
1.15643(55) & \hbox{if $d=3$}.
}
\label{26}
\end{equation}
If the analysis is done using the geometric average of the data instead,
then the estimate $\gamma=1.15662$ is obtained instead in the
cubic lattice.  

If the analysis is repeated, but now using the best estimates for 
$\log \mu_d$ (equation \Ref{bestmu}), then a two parameter fit using the model
in equation \Ref{eqn15} gives
\begin{equation}
\gamma = \cases{
1.34319(56) & \hbox{if $d=2$}; \cr
1.15681(17) & \hbox{if $d=3$}.
}
\label{27}
\end{equation}

These results should be compared to the exact value $\gamma=1.34375$ 
in two dimensions \cite{D86} and the estimate $\gamma=1.15698(34)$ 
\cite{S11} in three dimensions and are in both cases accurate to three 
decimal places (see also the estimate $1.15695300(95)$ \cite{C17} 
for a more accurate estimate in three dimensions). The differences from 
the exact value and the estimate in \cite{S11} are shown in brackets as 
an error term in equations \Ref{26} and \Ref{27}.

The parallel implementation of PERM (using the flatPERM implementation 
of PERM) in this paper makes it possible to exploit the parallel architecture 
of modern computers by feeding a flatPERM-sequence to each thread.  
Each sequence is recursively evolved by the algorithm and the exchange of 
information between sequences occurs by the use of shared data which 
incorporates information about the ensemble landscape from the other 
sequences into a given sequence, thereby affecting its future evolution.  
This approach can be used in the same way to implement a parallel 
GARM algorithm (see reference \cite{RJvR08}).  Closer integration of 
communication between parallel sequences may also be imagined; 
for example, two sequences sampling walks in the square lattice may
be considered as a single sequence sampling a path in the four 
dimensional hypercubic lattice.  This approach may also give accelerated 
convergence but a parallel implementation may not be possible, 
as the four dimensional path will have to be sampled in a single thread.   

We have also implemented an integrated parallel implementation of the 
Wang-Landau algorithm \cite{WL01} using a set of interacting sequences 
on state space similar to algorithm 2.  The Wang-Landau algorithm 
directly estimates the density of states by carrying out a random walk 
in energy space.  It tracks the energy $E$ of a system:  If the current 
energy $E_{old}$ ($g(E_{old})$) is the energy (respectively density) 
of the current configuration and $E_{new}$ ($g(E_{new})$) 
is the energy (respectively density) of the new proposed configuration, 
then the move is accepted with probability 
$p=\min\left\{\frac{g(E_{old})}{g(E_{new})},1\right\}$. 
Each time a state is visited by a sequence, the density of states 
is updated by a modification factor $f$ such that $g(E)\leftarrow g(E)\cdot f$. 
A histogram $H(E)$ of each visit is also kept and a 
flatness criterion for the histogram is used to update the modification 
factor $f$. That is, when the histogram achieves the flatness criterion 
it is reset and $f$ is reduced in a predetermined fashion. 
Care is usually taken here since if $f$ is decreased too rapidly this can lead 
to saturation errors (see reference \cite{belardinelli2007wang}).

The parallel implementation for the Wang-Landau algorithm differs slightly from 
that of the PERM algorithm and an earlier approach of Zhan \cite{zhan2008parallel}.
In our approach the parallel streams are used to control the update of a 
common $f$. The density of states for each stream are compared to estimate 
the error and then the updated $f$ value depends on this estimated error. 
That is, as the error declines the values of $f$ also decline.  The standard 
observed relationship is that the statistical error scales proportionally 
with $\sqrt{\log{f}}$ (see reference \cite{zhou2005understanding}).

The benefit of dynamically adjusting the parameter $f$ is that the $f$ values 
decline rapidly when the algorithm is converging quickly and vice versa. 
In particular, as in the case of the PERM algorithm, we find that the initial 
rate of convergence is significantly accelerated. Previous works have 
suggested that in the absence of additional information, an optimal convergence 
rate might be achieved by decreasing $\log f$ at a rate of $1/t$ where $t$ 
is the normalized time of the simulation \cite{zhou2008optimal}. Moreover, 
numerical results suggest that this achieves a statistical error of 
$1/\sqrt{t}$ and, in general, a theoretical upper bound on the error 
behaviour was shown in reference \cite{zhou2008optimal} to be $1/t$. 
By taking advantage of the additional information provided by the communicating 
sequences in our algorithm, we report that for reasonable length simulations 
the estimates of $c_n$ are found to greater accuracy than those from 
independent parallel implementations of the standard $1/t$ algorithm.

\vspace{1cm}
\noindent{\bf Acknowledgements:} EJJvR acknowledges financial support 
from NSERC (Canada) in the form of Discovery Grant RGPIN-2019-06303.  
SC acknowledges the support of NSERC (Canada) in the form of a Post 
Graduate Scholarship (Application No. PGSD3-535625-2019).

\vspace{2cm}
\noindent{\bf References}
\bibliographystyle{plain}
\bibliography{ParallelPerm}

\end{document}

%% file: macros.tex
\def\2;{\;\;}

\def\Pr{{\hbox{P}_{\hskip -1mm r}}}

\def\IntZ{{\mathbb Z}}

\def\mathS{{\mathbb S}}

\def\o#1{\overline{#1}}

\def\Ref#1{(\ref{#1})}

\def\C#1{{\mathcal #1}}

\def\sigmas#1{{\sigma_#1}}

\def\omegas#1{{\omega_#1}}

\def\Sfrac#1#2{\hbox{\large $\frac{#1}{#2}$}}
\def\sfrac#1#2{\hbox{\nor $\frac{#1}{#2}$}}

\def\LB{\left(}         \def\RB{\right)}
\def\LV{\left|}        \def\RV{\right|}
\def\LA{\left\langle}        \def\RA{\right\rangle}
      
\def\lcl{\left\lceil}  \def\rcl{\right\rceil}
\def\lfl{\!\left\lfloor} \def\rfl{\right\rfloor\!}
       
\def\LH{\left[}        \def\RH{\right]}

\def\nor{\normalsize}
\def\fns{\scriptsize}

\def\thin{ {\hspace{0.75pt}} }

\def\minus{{\hspace{0.85pt}{-}\hspace{0.85pt}}}

\def\tiny#1{\scalebox{0.75}{#1}}

%% file: macros-pictex.tex
\definecolor{blue}{rgb}{0,0.18,0.39}
\definecolor{RoyalBlue}{rgb}{0,0.2,0.7}

\definecolor{Maroon}{cmyk}{0,0.87,0.68,0.62}
\definecolor{Brown}{rgb}{0.7,0.3,0}
\definecolor{Navy}{rgb}{0.3,0.0,0.4}
\definecolor{Red}{cmyk}{0,1,1,0}
\definecolor{BrickRed}{cmyk}{0.16,0.89,0.61,0.02}
\definecolor{DarkRed}{cmyk}{0,1,1,0.70}
\definecolor{DarkBlue}{cmyk}{1,1,0,0.2}
\definecolor{DarkGreen}{cmyk}{1,0,1,0.4}
\definecolor{Green}{cmyk}{1,0,1,0}
\definecolor{DarkBrown}{cmyk}{0,0.81,1,0.6}
\definecolor{OrangeRed}{cmyk}{0,1,0.87,0}
\definecolor{RedOrange}{cmyk}{0,0.77,0.87,0}
\definecolor{Orange}{cmyk}{0,0.61,0.87,0}
\definecolor{Offwhite}{rgb}{.8,0.9,.8}
\definecolor{Offwhite2}{cmyk}{.04,.02,.01,0}
\definecolor{Tan}{rgb}{0.82,0.70,0.55}
\definecolor{Blue}{rgb}{0,0,1}
\definecolor{RoyalBlue}{rgb}{0.25,0.41,0.88}
\definecolor{Sepia}{rgb}{0.37,0.14,0.07}
\definecolor{myblue}{cmyk}{0.025,0.05,0,0}
\definecolor{Mahogany}{cmyk}{0.18,0.87,1,0.08}

\definecolor{green1}{cmyk}{0.25,0,0.76,0}
\definecolor{green2}{cmyk}{0.25,0,0.76,0.07}
\definecolor{green3}{cmyk}{0.25,0,0.76,0.20}
\definecolor{green4}{cmyk}{0.25,0,0.75,0.30}
\definecolor{green5}{cmyk}{0.25,0,0.75,0.40}
\definecolor{green6}{cmyk}{0.25,0,0.75,0.50}

\definecolor{B02}{cmyk}{0,0.14,0.22,0.12}
\definecolor{B03}{cmyk}{0,0.16,0.26,0.16}
\definecolor{B04}{cmyk}{0,0.19,0.28,0.19}
\definecolor{B05}{cmyk}{0,0.25,0.32,0.25}
\definecolor{B06}{cmyk}{0,0.31,0.36,0.31}
\definecolor{B07}{cmyk}{0,0.37,0.40,0.37}
\definecolor{B08}{cmyk}{0,0.46,0.46,0.46}
\definecolor{B09}{cmyk}{0,0.55,0.52,0.54}
\definecolor{B10}{cmyk}{0,0.69,0.61,0.62}
\definecolor{B11}{cmyk}{0,0.78,0.70,0.68}
\definecolor{B12}{cmyk}{0,0.93,0.85,0.60}
\definecolor{B13}{cmyk}{0.5,1,0.6,0.30}
\definecolor{B14}{cmyk}{1,1,0,0.30}
\definecolor{B15}{cmyk}{1,1,0,0}